\documentstyle[preprint,aps,epsf]{revtex}                  
\global\let\epsfloaded=Y 
\begin{document}
\pagestyle{empty}                                      
\preprint{
\font\fortssbx=cmssbx10 scaled \magstep2
\hbox to \hsize{
\hfill $
\vtop{
 \hbox{ }}$
}
}
\draft
\vfill
\title{Accelerating Universe and Event Horizon}
\vfill
\author{Xiao-Gang He}
\address{
\rm Department of Physics, National Taiwan University,
Taipei, Taiwan 10764, R.O.C.}

%
%
\vfill
\maketitle
\begin{abstract}
It has been argued in the literature 
that if a universe is expanding with an
accelerating rate indefinitely, it presents a challenge to string
theories due to the existence of event horizons. We study the
fate of a currently accelerating universe. We show
that the universe will continue to accelerate indefinitely if the
parameter $\omega = p/\rho$ of the equation of state is a
constant, no matter how many different types of energy (matter,
radiation, quintessence, cosmological constant and etc) are contained
in the universe.
This type of universe will always exhibit an event horizon
indicating that such a universe may not be derived from 
string theories. We also comment on some related issues.
\end{abstract}
%
%
\pacs{PACS numbers:
 }
%
%
\pagestyle{plain}

\noindent
{\bf Introduction}

It is well known that from observation of red-shift of light from
distant galaxies and other evidences that our universe is an expanding 
one\cite{1}.
Recently direct evidences from the study of Hubble diagram
for Type Ia supernovae\cite{2} show that our universe is not only expanding,
but also undergoing through an
accelerating expansion era. Also data from cosmic
microwave background (CMB) radiation\cite{3} indicate that the
dominant energy form of our universe may be from a cosmological
constant which also leads to an accelerating expansion. If the
accelerating expansion last forever the universe will exhibit an
event horizon, there exist regions of the universe which will be
inaccessible to light probes. It has been argued that a universe
exhibiting an event horizon presents a challenge for string
theories, which are the only known theories can treatment gravity and
quantum mechanics consistently, that among several difficulties it
is not possible to construct a conventional S-matrix because the
local observer inside his/her horizon is not able to isolate
particles to be scattered\cite{4,5}. Some specific examples of 
a universe exhibiting event horizons 
have been studied\cite{5,5a}. In this paper we study whether
a universe will undergo an accelerating expansion indefinitely if it is 
currently accelerating, for 
a class of models with several forms of energy. 

For a universe which contains 
different energies, there are
some subtle issues need to be handled with care.
To start with we emphasis that a universe with dominant energy
coming from cosmological constant if not carefully balanced, for
that matter any form of quintessence, may not be able to evolve
from the past to present. With a Friedmann-Robertson-Walker (FRW)
metric, 

\begin{eqnarray}
d^2s = d^2t - R^2(t)[ {d^2 r \over 1-k r^2} + r^2( d^2\theta + \sin^2\theta
d^2\phi)],
\end{eqnarray}
and conservation of energy, the energy density at different times, 
if each type of energy is separately coinserved, 
is related to the scaling factor
$R(t)$ by\cite{1,6}

\begin{eqnarray}
\Omega^i = \Omega^i_0 x^{-3(1+\omega_i)},\;\;\;\; x = {R\over R_0},
\end{eqnarray}
where $\Omega_i$ is the ratio of the energy density $\rho_i$
of the ith type of energy to the
critical energy density $\rho_c = 3H_0^2/8\pi G_N$ with $H_0$ being the
present day Hubble constant. The subscript
``0'' on the various quantities
indicate the values at the present time.

The age of the universe is given by

\begin{eqnarray}
t=\int^t_{0^+} dt = \int^R_{0^+}{dR\over \dot{ R}}\;.
\end{eqnarray}
Here $0^+$ denotes the earliest time where the FRW Big-Bang theory
can be applied. Before that one needs to take into account effects from 
inflation and quantum gravity.

Using the Friedmann equation\cite{1,6}

\begin{eqnarray}
H^2 = \left ( {\dot {R}\over R}\right )^2
&=& {8\pi G_N\over 3}\sum_i \rho_i -{\kappa\over R^2}\nonumber\\
&=& H^2_0 [\sum_i \Omega^i_0 x^{-3(1+\omega_i)} + (1-\sum_i \Omega_0^i)x^{-2}],
\end{eqnarray}
one obtains

\begin{eqnarray}
t ={1\over H_0} \int^{x}_{0^+} {dy\over \sqrt{\sum_i \Omega_0^i (y^{-(1+3\omega_i)}
-1) +1}}.
\end{eqnarray}
In the above we
have used the identity
$\Omega^\kappa = -3\kappa/8\pi G_N R^2 H^2
= 1-\sum_i\Omega^i$.

The above equation is only meaningful when

\begin{eqnarray}
f(x) = \sum_i \Omega_0^i (x^{-(1+3\omega_i)} -1) +1  = \sum_i
\Omega_0^i x^{-(1+3\omega_i)} + \Omega^\kappa
> 0.
\end{eqnarray}
This is automatically true for a universe with $\kappa =0$ and
$\kappa = -1$. For $\kappa = 1$ it is not guaranteed to have
$f(x)>0$. For illustration
we apply the above formula to a hypothetical universe with
only $\Omega^m_0$ and $\Omega_0^\Lambda$. We find

\begin{eqnarray}
f(x) = \Omega_0^m x^{-1} + \Omega_0^\Lambda x^2 +
(1-\Omega_0^m-\Omega^\Lambda_0).
\end{eqnarray}

With $\Omega_0^m = 0.1$ and $\Omega^\Lambda_0 = 1.5$, the
universe is an accelerating one at present 
with dominant energy from cosmological constant. 
However this is a universe which can not evolve from the past to the 
present because $f(x)$ is not always
larger than zero. 
We find that there are three roots for $f(x)$
at $x$ equal to: -0.70, 0.18, and 0.51. The regions with $f(x)
>0$ are: $-0.70 < x < 0.18$, and $x
>0.51$. This implies that if the universe started from $x =0$, it
can only evolve to $x=0.18$ with a finite life time, but can not evolve
to the present ($x=1$); Or it can only evolve from $x=0.51$ to the 
present and to the
future indefinitely. To have a physical universe which can evolve
from $t=0^+$ to the present and have a future, 
the energy distributions are constrained.
One should be careful to select cosmological parameters to make
sure that the universe is physical in all times, past, present and
may be future.
\\
\\
\noindent
{\bf Accelerating Expansion}

In order to have an accelerating expansion, one needs to have
the deceleration parameter $q =-\ddot{ R} /H^2_0R$ to be less than zero.
From Einstein equation\cite{1,6}, 

\begin{eqnarray}
{\ddot{R}\over R} = - {4\pi G_N\over 3} (\rho +3 p),
\end{eqnarray}
one can write $q$ as

\begin{eqnarray}
q(x) = {1\over 2} \sum_i\Omega^i_0 x^{-3(1+\omega_i)} (1+3\omega_i).
\end{eqnarray}
An accelerating universe at present time requires the present deceleration
parameter $q_0$ to be less than 0, that is

\begin{eqnarray}
q_0=q(1)= {1\over 2} \sum_i\Omega^i_0 (1+3\omega_i) < 0\;,
\end{eqnarray}
which implies that at least one of the energies has $\omega_i
<-1/3$. Energy forms with $\omega <-1/3$ have been studied
extensively in the literature from quintessence to cosmological
constant\cite{1,7}. Quintessence induced by scalar potentials
has a $\omega$ in the range
of $-1$ to 1, while cosmological constant has a
$\omega = -1$.

If the present universe contains only one type of energy
and this energy has a constant $\omega < -1/3$, the universe is 
guaranteed to be in
accelerating expansion state and will  do so forever in the future. It
should be noted that this universe can not be a closed one because 
in the distant past with $x\to {0^+}$, $f(x) <0$ as have been discussed 
in the
previous section. To describe a universe for all times, there must
be other forms of energy with $\omega > -1/3$. However if the
universe contains more than one types of energy, one of the
energies with $\omega < -1/3$ may not be enough to have an
accelerating expansion at present. Let us consider a universe which
contains just two types of energy with one of them having $1+3\omega_1 >
0$ and another having $1+3\omega_2 <0$. The accelerating expansion
requires

\begin{eqnarray}
\omega_2 < -{1\over 3} [1 + (1+3\omega_1){\Omega_0^1\over \Omega^2_0}].
\end{eqnarray}
 For a universe with a matter energy $\Omega_0^m = 0.3$ and a quintessence
energy
$\Omega_0^q = 0.7$, one would have to have $\omega_q < -10/21$ which is
considerably smaller than $-1/3$.

It is also interesting to note that with the constraint 
$\omega_2 >-1$, an accelerating
expansion implies

\begin{eqnarray}
 {\Omega^1_0\over \Omega_0^2} < {2\over 1+3\omega_1}.
\end{eqnarray}
For a universe with matter and one quintessence,
$\Omega^m_0/\Omega^q_0$ must be less than 2.

We now study whether a universe will forever accelerate 
if it is presently accelerating with many
different types of energy. The answer to this problem 
depends on how the parameter $\omega_i$
for each form of energy changes with time. Without detailed
information on this, it is not possible to obtain an answer. Here
we will consider a very simple and interesting case
that all $\omega_i$ do not change with time. In this case the answer is
simple that a currently accelerating universe will maintain its
acceleration forever, no matter how many different
types of energy are contained in the universe. Of course to make
sure the universe can evolve from $t=0^+$ to the present, we have to also 
assume $f(x)>0$ for all $x > 0^+$ is satisfied. 

The decelerating parameter can be written as a
sum of the contribution ($A(x)$) from all energies with $1+3\omega_i >0$, and
of the contribution ($B(x)$) from all energies with
$1+3\omega_i <0$, as

\begin{eqnarray}
&&q(x) = {1\over 2}{1\over x^2}[A(x) + B(x)],\nonumber\\
&&A(x) =  \sum_{1+3\omega_i>0}
\Omega^i_0 x^{-(1+3\omega_i)}(1+3
\omega_i),\nonumber\\
&&B(x) = 
\sum_{1+3\omega_i<0}\Omega^i_0 x^{-(1+3\omega_i)}(1+3\omega_i).
\end{eqnarray}
By definition,  $A(x) >0$ and $B(x)<0$.

The condition $q_0 <0$ requires $A(1) < |B(1)|$. Since $A(x)$
is composed of terms with negative powers in $x$ and $B(x)$ is composed of
terms with positive powers in $x$, it is clear that

\begin{eqnarray}
A(x>1) < A(1),\;\;\;\;|B(x>1)| > |B(1)|.
\end{eqnarray}
This leads to

\begin{eqnarray}
q(x>1) < q_0 < 0.
\end{eqnarray}
The universe will accelerate forever.

In the past $q(x) <0$ is not guaranteed. For small
enough $x$, $A(x)$ will become larger than $|B(x)|$. The universe
must had gone through a decelerating era and evolved to the present 
accelerating era. For example, if there is only one dark energy with 
$\omega^q=p/\rho$, the
transition of deceleration and acceleration occured at
$x=(-(1+3\omega^q)\Omega_0^q/\Omega^m_0)^{1/3\omega^q}$.
With $\Omega^m_0 = 0.3$ and $\Omega_0^q=0.7$, for vacuum energy the
transition happened at redshipt $z=0.67$. This is consistent with the
recent data from observations of SN 1997ff at $z\sim 1.7$\cite{red}. 
\\
\\
\noindent
{\bf Existence of Event Horizons}

If the universe is accelerating indefinitely, for any two objects
separated by a fixed comoving distance, their relative proper
speed will reach the speed of light after sometime, and they will
cease to communicate. Therefore the universe will exhibit an event
horizon. More precisely if the integral

\begin{eqnarray}
D_H(\infty) = \int^\infty_{t'} {dt\over R(t)},
\end{eqnarray}
is finite,
the universe exhibits an event horizon\cite{5}.

We now show that under the conditions given in the previous
section, $D_H(\infty)$ is finite. To this end let us first
consider the behavior of $R(t)$ at large t. Since the universe is
accelerating, at least one of the $\omega_i$ is less than
$-1/3$, and some of the terms in $f(x)$ defined in eq. (6) must
have positive powers in $x$. Letting $\omega_l$ be the smallest one 
of all $\omega_i$,
at sufficiently large t and $t'$, $f(x) \approx \Omega_0^l
x^{-(1+3\omega_l)}$. One has

\begin{eqnarray}
t-t' = {1\over H_0} \int^{R/R_0}_{R'/R_0} {dx\over \sqrt{f(x)}}
\approx   {1\over H_0 \Omega_0^{l\;1/2}}\int^{R/R_0}_{R'/R_0} 
x^{(1+3\omega_l)/2} dx.
\end{eqnarray}
and therefore,

\begin{eqnarray}
R(t) &\sim& t^{2/3(1+\omega_l)},\;\;\;\;\;\;\mbox{for}\;\; 
-1 < \omega_l < -1/3,
\nonumber\\
&\sim& e^{\Omega_0^{l\;1/2} H_0 t},\;\;\;\;\;\;\mbox{for}\;\; \omega_l = -1.
\end{eqnarray}

Using the above, one obtains

\begin{eqnarray}
D_H(\infty) &\sim& \int^{\infty}_{t'} t^{-2/3(1+\omega_l)} dt \sim
t^{(1+3\omega_l)/3(1+\omega_l)} |_{t=\infty}^{t'},\;\;
\mbox{for}\;\;-1< \omega_l < -1/3,\nonumber\\ &\sim&
\int^{\infty}_{t'} e^{-\Omega_0^{l\;1/2}H_0 t} dt,\;\;\;\;\hspace{3.7cm}
\mbox{for}\;\; \omega_l = -1.
\end{eqnarray}
It is obvious that for the case with 
$\omega_l=-1$, $D_H(\infty)$ is finite. For
the case with  $-1< \omega_l <-1/3$, since $(1+3\omega_l)/3(1+\omega_l)
<0$, $D_H(\infty)$ is also finite.

$D_H(\infty)$ is finite can also be shown by rewriting 
$D_H$ in terms of $\Omega_0^i$, $\omega_i$ and $x$ as

\begin{eqnarray}
D_H(\infty) &=&\int^\infty_{R'} {dR\over R \dot{R}} = {1\over R_0}
\int^\infty_{x'} {dx\over x^2 H}
\nonumber\\
&=& {1\over R_0H_0}\int^\infty_{x'} {dx \over x \sqrt{\sum_i
\Omega_0^i(x^{-(1+3\omega_l)}-1) +1}}.
\end{eqnarray}
For large enough $x'$, $D_H$ can be well approximated by

\begin{eqnarray}
D_H(\infty) \approx {1\over R_0H_0 \Omega_0^{l\;1/2}}\int^\infty_{x'}
x^{(1+3\omega_l)/2 - 1} dx.
\end{eqnarray}
Since $1+3\omega_l <0$, the above integral is finite.

We therefore have shown that a universe always exhibits an event
horizon if it is accelerating at present with
constant $\omega_i$.
\\
\\
\noindent
{\bf Discussions}

There are indications that the matter (visible,
baryonic and dark matter) contributes about one third of the
energy, $\Omega_0^m \approx 0.3$\cite{1}. 
Also from CMB data\cite{3}, the total
energy is determined to be very close to one, if the rest of the
energy is from one type of energy (neglecting small radiation
energy), then $\Omega_0^q \approx 0.7$. 
If the evidences from type Ia supernovae\cite{2} showing that the
present universe is accelerating are confirmed, it is possible that
70\% of the energy in our universe is composed of  energies 
with $\omega_q <-1/3$. If the universe contains just matter and one other 
form of energy, its $\omega_q$ must be less than $ -0.4762$. 
This energy could be due to cosmological
constant or quintessence. The universe may forever accelerate 
and exhibit an event horizon. 
This may present a challenge
to string theories which are candidate theories describing
consistently gravity and quantum mechanics, and
therefore the cosmology\cite{4,5}. The accelerating universe discussed in
the previous sections may not be obtained from string theories.

The conclusion that a presently accelerating universe will lead to
the existence of event horizons crucially depends on the
assumption that the parameters $\omega_i$ for the equation of states
are constant in time. If $\omega_i$ changes with time, the situation
may change. 
Several calculations based on quintessence models with varying $\omega_q$  
concluded that a currently accelerating universe
will continue to accelerate and exhibit event horizons\cite{4,7}. 
In fact from our previous discussions 
that even if $\omega_i$ change with time 
and the universe is not accelerating now as long as one of the $\omega_i$ 
will be less than $-1/3$ from some later time on, the universe will 
eventually become an accelerating 
one and exhibit an event horizon. 

Our universe,
of course, may be much
more complicated than what have been assumed in the previous sections and 
more complicated than some of the simple quintessence models. 
One can not exclude
the possibility that the presently
accelerating universe will become a decelerating one or a universe with
$q>-1/3$ in the future due to some mechanisms. In fact such models exist such
as the mechanism discussed in Ref.[\cite{8}] that for a class 
of scalar potentials the vacuum domination is only a transient phemonenon.
Quintessence scalar decays
due to interactions with matter can also stop the acceleration\cite{9}. 
To have a better understanding of
the fate of our universe, accelerating forever or not, more experimental 
information about present day cosmological 
parameters and further theoretical studies are needed.

I thank T.H. Chiueh and Miao Li for discussions. This was supported in part
by National Science Council under grant NSC 89-2112-M-002-058 and in part
by the Ministry of Education Academic Excellence Project 89-N-FA01-1-4-3.

\end{document}